\documentstyle[preprint,prd,aps,epsf]{revtex}
\begin{document}
\title{\bf A Renormalization Group for Dynamical\\
           Triangulations in Arbitrary Dimensions }
\author{\bf Ray L. Renken }
\address{ Department of Physics, University of Central Florida,
Orlando, Florida, 32816 }
\date{July 29, 1996}
\maketitle
\mediumtext
\widetext
\begin{abstract}
\begin{quotation}

A block spin renormalization group approach is proposed for the dynamical
triangulation formulation of quantum gravity in arbitrary dimensions.
Renormalization group flow diagrams are presented for the three-dimensional
and four-dimensional theories near their respective transitions.

\end{quotation}
\vskip .25in
\end{abstract}
\section{ Introduction }
There are currently many different avenues to a theory of quantum gravity.
This paper focuses on the formulation provided by dynamical
triangulations.  Extensive work has
been done in two dimensions and a significant body of work has also examined
three and four dimensions \cite{dtreview}.  Dynamical triangulations are
similar to the formulation
of Regge in that the manifold is broken up into simplexes and the curvature
is identified with a deficit angle about a subsimplex with dimensionality two
less than that of the manifold.  The formulations differ in their treatment of
the functional integral over all possible metrics.  In the Regge approach, this
is provided by varying the link lengths on a fixed lattice.  In the dynamical
triangulation approach, the link lengths are held fixed but the lattice is
varied: all possible triangulations are summed over.  Monte Carlo evaluations
of the sum are performed by randomly wandering through the space of
triangulations.  It has been proven that any triangulation can be reached from
any other triangulation (for a given topology) through a sequence of local
moves \cite{migdal}.  A simple algorithm
allows the computation of functional integrals over metrics in any desired
dimension with a computer program only 350 lines long \cite{catter}.

The simplest observables in these models are the number of subsimplexes of
various dimensions (e.g. nodes, links, triangles, etc.) and these numbers
can be used to construct an action.  They are related due to
topological constraints and the requirement that the triangulations yield a
manifold: in three and four dimensions the number of nodes and the number of
simplexes determine all of the other numbers of subsimplexes.
The basic action used in these theories is
\begin{equation}
S = \alpha N_0 - \beta N_D \label{action}
\end{equation}
where $\alpha$ and $\beta$ are chemical potentials corresponding to Newton's
constant and the cosmological constant, respectively, while $N_0$ is the number
of nodes and $N_D$ is the volume.  In two dimensions, $N_0$ and $N_D$
are related so there is only one term in the action.
In fact, it is possible to work in an ensemble where the volume is fixed
(leaving zero terms in the action).  This is possible because ergodicity has
been proven for a restricted set of update moves that do not alter the volume.
In three and higher dimensions there is no corresponding proof: the full
ensemble including volume fluctuations must be used \cite{gross}.

To recover a viable continuum theory from the lattice theory requires two
limits to be taken.  First, there is the thermodynamic, large volume,
limit.  It is reached by adjusting $\beta$.  When $\beta$ is large, the
volume is small.  As $\beta$ is decreased, the volume increases until a
critical value of $\beta$ is reached at which the volume diverges.  In
practical calculations, $\beta$ must be taken slightly larger than the
critical value and there are finite size effects.  The other limit that must
be taken is the continuum
limit.  A non-zero correlation length occurs in the continuum, where the
lattice spacing is taken to zero, only when the correlation length in lattice
units is infinite, so there must be a second order phase transition
present in the phase diagram of the theory.  
In spin theories and lattice gauge theories, the block spin renormalization
group approach is a natural tool with which to study continuous phase
transitions \cite{rg1,rg2,rg3,rg4,rg5,rg6,rg7}.  This paper develops a
renormalization group approach for dynamical triangulations.
It is used here to examine the flow of couplings, but it also provides for the
systematic calculation of critical couplings and critical exponents if a second
order transition can be found.

As an excellent example of the potential of the renormalization group approach
see the application of it to the three-dimensional Ising model in \cite{ising}.
Before applying this idea to the lattice quantum gravity models, it is
useful to consider the differences between the Ising model and a dynamically
triangulated model.  In the former there is a fixed lattice, of linear dimension
$L$, on which the degrees of freedom are arranged in a regular way.  It is
simple to draw a courser (block) lattice, of linear dimension $L/b$, as a subset
of the original one which partitions the spins into identically structured
groups (such as $b$ by $b$ blocks).  These
groups can then be averaged according to a specified rule in order to convert
the original spin configuration $C$, composed of $L^D$ spins, into a block spin
configuration $C^{\prime}$, composed of $(L/b)^D$ spins.  In contrast, in
a dynamically triangulated model the lattice is not fixed and the
lattice structure itself embodies the degrees of freedom.

Physically, triangulations are interpreted as embodying information about the
metric structure of spacetime: points connected by a link are closer than
points not connected by a link.  Presumably this should be true for block
triangulations as well and, in two dimensions, the condition can be
imposed easily to define a block spin renormalization group
transformation \cite{renken,rck}.  This is accomplished as follows.
In two dimensions at fixed volume the set of local update moves is a single
move, the link flip.  View a link and the two triangles that share it as a
square with one diagonal drawn.  A link flip simply replaces the diagonal
with the alternative one.  It is the simplicity of the link flipping move that
makes it easy to define the block triangulations.  Beginning with the original
triangulation, $T(N_0,N_D)$, illustrated with thin lines in Fig. 1, the
$N_0^{\prime}$ block nodes are defined as some subset of the original nodes.
A block triangulation, $T^{\prime}(N_0^{\prime},N_D^{\prime})$, illustrated
with thick lines in the figure, is required to have the property that the
minimum number of links on the original lattice required to travel from one
block node to another block node connected to it by a block link must be
smaller than the corresponding number of links that must be traversed to
travel between the block nodes associated with the alternative diagonal.
Block links are flipped until this requirement is satisfied.  Once an
initial block triangulation has been obtained, updates of the original lattice
are followed by updates of the block lattice so as to maintain the rule.
The simplest way to obtain an initial block triangulation is to start with
a regular original triangulation where it is easy to generate a course lattice
with the required property.  Matter can also be incorporated into this scheme.
This renormalization group transformation gives quite good results for coupling
constant flows and critical couplings, but does not do well on critical
exponents.

A variety of subsequent renormalization group transformations have been
proposed \cite{johnston,thorli}.  The most successful is the node deletion
blocking scheme which is very simple and gives excellent results \cite{thorli}.
The idea, still working in two dimensions, is
to start with an original triangulation with $N_0$ nodes and $N_2$ triangles
($N_2 = 2(N_0 - \chi)$ in two dimensions with $\chi$ the euler characteristic
of the surface) and to delete $R$ nodes in order to produce a block
triangulation with $N_0^{\prime} =
N_0 - R$ nodes and $N_2^{\prime} = N_2 - 2R$
triangles.  Consider the deletion of a single particular node labeled $P$.
The node $P$ belongs to a number of triangles forming a set to be referred to
as $B$.  The set $B$ forms a two-dimensional ball around the node $P$.  The
way $P$ is eliminated is by removing the entire ball $B$ from the manifold
and then filling the resulting hole back in with new triangles, but without
any new nodes.  The following algorithm is used to accomplish this.  A link
attached to $P$ is
chosen at random and, if it is an allowed move, flipped.  As a result
two old triangles are removed from the set $B$ and one new one is introduced,
decreasing the total by one.  Links are repeatedly randomly chosen and flipped
until there are only three triangles remaining in the ball.  Then the node
and the remaining three triangles are replaced with one new triangle.  The
original ball has been completely removed and the hole has been refilled.
Fig.~2 schematically represents the procedure.
$R$ randomly chosen nodes are deleted in this way and the resulting
lattice is interpreted as resulting from a block spin renormalization group
transformation.  When the standard renormalization group formalism is used to
extract critical couplings and exponents for the two-dimensional Ising model
coupled to dynamical triangulations, the results agree with the known exact
solutions.

Since the node deletion procedure works so well in two dimensions, it is
natural to try to apply it in higher dimensions.
As mentioned earlier, it is possible to formulate and implement the
update moves for triangulations in such a way that the dimensionality of the
simplexes is an easily modified variable.  The purpose of this paper is to
put the node deletion scheme in this more general setting.  Obviously, it is
simple to identify and eliminate all of the simplexes associated with a given
node regardless of the dimension.  The tricky part is filling the hole back in.
The reason two dimensions at fixed volume is different from a general
dimension,
$D$, is that the former has only one update move (the link flip) while the
latter, using the formulation in \cite{catter}, has $D + 1$.  If $i$ ranges from
$0$ to $D$, then there is a move, for each $i$, that replaces $D+1-i$ simplexes
with $i+1$ new ones.  $i$ is the dimension of a subsimplex that must be
chosen to make the move.  For instance, a link flip has $D=2$ and $i=1$ which
means a one-dimensional subsimplex (the link) must be chosen to make the
move and two old simplexes are replaced with two new ones.  In two dimensions,
the simplexes associated with a chosen node are eliminated by making link flips
which remove simplexes from the node's association until the minimal number
are left. The node is then eliminated.  This final step is actually an
$i=0$ move.  In higher dimensions, the set of simplexes associated
with a node can be reduced using the $i > 0$ moves until the minimal number
is reached and a final $i=0$ moves completes the operation.  By setting up
the problem in this way, it is possible to write a program that does node
deletion for a simulation of dynamical triangulations in any dimension.
Changing dimensions requires the alteration of a single character in the
program.

\section{ Details }

This section gives details which can be skipped, but which may be helpful
to those who wish to implement the scheme.  To produce a block configuration,
start by copying the original configuration, along with all relevant
bookkeeping, into
new data structures.  The update routines described in \cite{catter} can be
modified so that they take arguments specifying which copy of the lattice
information is to be changed.  In this way, exactly the same routines
can be used to update the original lattice and to eliminate nodes in the blocked
lattice.  Aside from being convenient, this is a good way to avoid errors.
In the case of the block lattice, the update moves are not used to generate
configurations distributed in accordance with the action but to fill a hole
in the manifold made by eliminating a node and its associated simplexes.
Create a list of all of the simplexes associated with the node
that is to be eliminated.  These simplexes form a ball, $B$.  Just as in two
dimensions where links attached to the node are chosen at random and flipped
(when consistent with the geometrical constraints) so as to reduce the number
of associated simplexes, in arbitrary dimensions interior $i$-subsimplexes are
chosen at random and the update move labeled by $i$ is made.  $i$ cannot be zero
until the node itself is finally removed.

The interior subsimplex in question is chosen by choosing one of the simplexes
in the
list at random and then choosing $i+1$ of its nodes at random.  The other
$D-i$ nodes specify $D-i$ neighboring simplexes because for a marked node
in a given simplex
there is one other simplex that shares all of the nodes in the given simplex
except for the marked one.  These neighbors play a role in the algorithm of
\cite{catter}.  Each of the $D-i$ neighboring simplexes has a new node not
found in the given simplex and
all $D-i$ of these new nodes must be the same if the contemplated move is an
allowed one.  If the contemplated move passes this test it is referred
to as ``legal".  The $D-i$ neighbors play an additional role here in that they
must all be in the list of simplexes associated with the node.  If they are
not, the chosen subsimplex is not interior to the ball.
Altering it would therefore not fill the hole, but would alter its boundary.

Until the last (node eliminating) move, $i$ can be any number from $1$ to $D$.
In two dimensions at fixed volume, where $i=1$ is the only possibility, there
is no choice
to be made.  In general $D$, many different algorithms for choosing $i$
are possible; this is just the freedom to choose different kernels in the
renormalization group transformation.  Some may be apt and some may not.
Only a single choice for the kernel is explored in this paper.  It is intended
to eliminate the ball of simplexes as rapidly as possible.  Since an $i$ move
removes $D-2i$ simplexes, it is likely that the quickest way to remove the ball
is to choose moves with $i$ as small as possible.  This is not quite as
obvious as it seems because, when a move is made in the interior of the ball,
additional simplexes are lost: some of the new ones are no
longer associated with the node and therefore do not belong in $B$.  $i=0$ is
not an option until the
last move, so $i=1$ is probably the best move.  However, sometimes no $i=1$
move is allowable, anywhere in the ball, because of
pre-existing connections on the surface of the ball due to exterior
simplexes.  In fact, sometimes the only possible move is node insertion, $i=D$.
Consequently, the
algorithm used here looks for and tries all possible $i=1$ moves,
then does the same for all possible $i=2$ moves and so on up to $i=D-1$.  If
no such moves are possible, a random simplex is chosen from the ball and a
node is inserted.  After each interior update is made, membership in the ball
(defined by having the targeted node as a vertex) is re-examined.
Then another interior update is made.  This is repeated until the minimal
number ($D+1$) of simplexes are associated with the node which is then
removed with an $i=0$ move.

For each move type, all of the simplexes (in random order) are checked for
appropriate possibilities.  This is done using the fact that a legal move
of type $i$ involves $D-i$ neighbors that share the same new node.  For
a chosen simplex in the ball, all of the new nodes from its neighbors 
(i.e. those conjugate to nodes in the chosen simplex) that are in the ball
are placed in a list which is checked for multiplicity.  Each new
node with the proper multiplicity correspond to a possible move.
These possible moves are drawn from at random and checked to see if the
geometrical constraints are satisfied.  If so, the move is made.  If not,
the list of possible moves is drawn from again.  If the list is empty then
no move at the given value of $i$ is possible and the procedure must be
repeated for the next value of $i$.

Instead of removing a node and its associated simplexes, one might consider
removing a link and its associated simplexes or, generally, an $X$-dimensional
subsimplex and its associated simplexes.  In two dimensions, a node is
associated with the curvature.  In general, a $(D-2)$-subsimplex is associated
with the curvature.  Perhaps the appropriate generalization of removing a node
in two dimensions is eliminating a $(D-2)$-subsimplex in $D$ dimensions.
If the subsimplex chosen for deletion is of dimension $X$ rather than zero,
essentially the same procedure as above can be used to fill the hole generated
by removing all of the simplexes associated with it.  The most important
difference is that now only moves of $i>X$ are allowed until the very last move
when $i=X$ is allowed.  In particular, if $X=D-2$, i.e. subsimplexes
associated with the curvature are eliminated, then the lowest
allowed value of $i$ is $D-1$ until the last move.  $i = D-1$ adds
$D-2$ simplexes each move.  In two dimensions the volume change is
zero which is consistent with the ensemble used there.  In higher dimensions
the volume increases.  One generally expects a renormalization
group transformation to produce a smaller system, one with fewer degrees of
freedom, so more thought would be necessary to interpret the results in
the case of $X=D-2$ with $D > 2$.  Even $X=1$ empirically appears to produce
larger volumes rather than smaller ones in three and four dimensions.
Throughout the rest of the paper, $X$ will always be zero.

While deletion of one node is interpreted as a renormalization
group transformation, it is generally useful to delete a number of nodes.
Perhaps the most naive way to choose the node is
to choose a simplex at random and then one of its nodes at random.  However,
the node chosen in this way is chosen with a probability proportional
to its ``order", defined to be the number of simplexes that contain it.
Choosing different
nodes with different probabilities could cause unwanted distortions
in the effective theory.  One remedy is to include a step which only
accepts the node with probability equal to the inverse of its order.
In this paper, since a large number of deletions are
typically used, a list of all of the nodes is constructed and chosen
from at random.  A brief study suggested that these alternatives have very
similar characteristics, both in speed and in results.

Note that two dimensions
is special in that the number of nodes and the number of $D$-simplexes
(triangles) is related: deleting a node requires the deletion of exactly
two triangles.  This is useful, because in the block spin renormalization group
approach there is typically a small target volume which one wishes to block to.
Comparing expectation values only when they are produced on the same size
lattice eliminates finite size effects.  For instance, if a system of
$5000$ simplexes is blocked down to $1000$ simplexes and the expectation values
of that blocked theory are compared to those of a system of $10000$ simplexes
blocked down to $1000$ simplexes, any differences in expectation values of
operators are not due to finite size
effects (the two systems are the same size) but to differences in the effective
theories produced in the two cases.  It is easy to see that $2000$ nodes must
be eliminated in the first case and $4500$ in the second.

In higher dimensions, the relationship between the number of nodes and the
number of simplexes is dynamically determined.  Eliminating all of the 
simplexes associated with a given node can delete anywhere between $D+1$ 
simplexes and a number of them comparable to the volume.  To target the
volume, it would be necessary to perform a sequence of runs, adjusting the
number of deleted nodes each time until finally reaching the desired volume.
It is more convenient instead to target the node number.  For a given coupling,
the number of nodes at various volumes is either known from previous work or
can be obtained with a single run.  Taking the difference between the starting
number of nodes and the target number of nodes gives a very good
estimate of the number of nodes that must be deleted from the larger lattice
to produce the same number as the smaller lattice.  A minor adjustment is often
still necessary, since node insertion moves are occasionally needed in the
process of eliminating the ball of simplexes associated with a deleted node.

Alternatively, instead of deleting a fixed number of nodes, nodes can be
deleted until a target node number is reached.  The number of deleted nodes
then becomes a fluctuating quantity while the final number of blocked nodes
is fixed.  A very brief study found that these two schemes give identical
answers.

Regardless of which scheme is used, node numbers can typically
be matched only to an accuracy of $0.5$, since the number of eliminated nodes
is an integer while the initial average number of nodes and the target number
of nodes are not. This can be patched with an interpolation of data from two
runs, or by noting that the node density is nearly constant over such small
changes in the node number.  Consistent results are obtained in all cases.  

\section{ Results }
Node deletion in two dimensions at fixed volume has been studied in
\cite{thorli}.  For pure gravity, the string susceptibility exponent and the
Hausdorff dimension were calculated.  Flow toward a fixed point was
verified using operators built from the local curvature and a measure term
was added and shown to be irrelevant.
For Ising matter, a fixed point was obtained at the (analytically
known) critical coupling and a variety of exponents (the string susceptibility
exponent, the magnetic exponent, and the thermal exponent) were correctly
computed.  These results demonstrate that the node deletion renormalization
group transformation is quite apt in two dimensions.

In two dimensions, once the
infinite volume limit is obtained by adjusting the cosmological constant,
the theory is also critical.  In higher dimensions an additional parameter
(at least) must be tuned.  Given the action, eqn. (\ref{action}),
and tuning the cosmological constant to give infinite volume, the only variable
left with
which to search for a continuous phase transition, and therefore a continuum
limit, is $\alpha$.  In both three and four
dimensions there is a phase transition as a function of $\alpha$ from a phase
where the ratio of the number of nodes to the volume goes to zero as the
volume increases to a phase where the same ratio approaches a non-zero
constant.  In
three dimensions the phase transition is first order while in four dimensions
the phase transition has been thought to be second order although now there
is some evidence that it is first order \cite{bialas,bakker}.
The renormalization group procedure detailed in the last section allows us
to study the renormalization group flow of these higher-dimensional theories.
Two operators are used to study the renormalization group flows:
the volume, $N_D$, and the logarithm of the order of the nodes, $M$, defined as
\begin{equation}
M = \sum_{i \in N_0} \ln \left( { O_i \over D + 1 } \right).
\end{equation}
The latter operator is associated with a measure term in the sense that if
a term of the form $\mu M$ is added to the action, it corresponds to a
continuum measure of the form
\begin{equation}
\prod_x g^{\mu / 2}
\end{equation}
where $g$ is the determinant of the metric \cite{brugmann}.

The target number of nodes used here and in the following sections is 
the number of nodes associated with an unblocked system with $300$
simplexes.  Smaller systems have the problem that when a large system is
blocked down to that size using a fixed number of node deletions the large
fluctuations in the block node number
occasionally reach zero.

Fig. \ref{fourd} shows the expectation values of the two
operators as a function of the degree of blocking for various couplings.
All of the
flows begin in the lower left hand corner, where $<N_4> = 300$.
Subsequent points in the plot
correspond to systems with initial volumes of $1000$, $2000$, $4000$, and
$8000$ simplexes, all of which have been blocked down to the indicated
number of nodes.  Lines have been drawn
through points sharing a given initial value of $\alpha$.  The values of
$\alpha$ are (from right to left) $2.3, 2.4, 2.5, 2.6$, and $2.7$ and are
chosen to span the transition region.
All runs were for at least a million
sweeps each.  Errors in the blocked volume, as a function of initial volume,
are typically 0.05, 0.8, 1, 3, and 4, respectively, with somewhat smaller
errors for large $\alpha$ and larger errors for small $\alpha$.  The
corresponding errors in $<M>$ are of order 0.05, 0.2, 0.4, 0.4, and 0.6.

Although all of the flows initially head toward the same vicinity,
they then split up.
For $\alpha = 2.3, 2.4$ and $2.5$, the volume quickly increases with the
number of deleted nodes, meaning that the node density quickly flows to
zero as the renormalization group transformation is iterated.  This is not
surprising, since it is the expected large distance behavior in the 
crumpled phase.  For $\alpha = 2.7$ the flow turns toward smaller volumes
although it is not clear how far it will go: for a given number of nodes,
the volume has a lower bound.  For $\alpha = 2.6$, the volume stays constant
for many iterations of the renormalization group, but continues to flow
slowly toward increasing $M$.  The slow flow near the end of the $\alpha = 2.6$
line suggests that a fixed point
is nearby and could be reached with a small adjustment in the action.  If
$\alpha$ is ruled to be insufficient, $\mu$ may be the appropriate parameter
to vary.  The renormalization group formalism itself can determine the
adjustments necessary in $\alpha$ and $\mu$ to obtain a fixed point, but
much higher statistics will be necessary to get a useful result (the last two
points on the $\alpha = 2.6$ line are only 1.5 standard deviations apart).

In four dimensions the peak in the node susceptibility with 8000 simplexes
(which moves toward greater $\alpha$ as the volume is increased) occurs
around $\alpha = 2.3$ \cite{ckr4d}. More recent estimates include transition
couplings in the range of $2.586(8)$ to $2.654(26)$ depending on assumptions
\cite{bialas}.  The results in the figure are compatible with these estimates.

Since the first order
character of the transition is much more firmly established in the case of
the three-dimensional dynamical triangulation model, it is interesting
to look at the renormalization group flows for this model and to see how
they compare to the four-dimensional one.
Fig. \ref{threed} shows the renormalization group flow lines for the
three-dimensional dynamical triangulation model for six values of $\alpha$:
$3.8, 3.9, 3.95, 4.0, 4.1,$ and $4.2$ (labeling the flows from right to left)
spanning the
transition region.  The errors, as a function of initial volume, are typically
0.1, 0.4, 0.8, 1.0, and 1.2 for the blocked volume and 0.04, 0.1, 0.4, 0.7,
and 0.7 for $<M>$.
The flows for the first two values of $\alpha$
flow strongly toward large volumes (and therefore small node densities) as
is expected in the crumpled phase.  Higher values of $\alpha$ flow toward
smaller volumes, similar to the smooth phase behavior in four dimensions.
There is no intermediate case for the chosen set of initial couplings and
therefore no indication of a nearby fixed point.  This
is consistent with the strongly first order character of the transition.

Finally, we consider the addition of a measure term to the action.
In \cite{brugmann}, they
suggest that the scaling behavior at the transition changes with the presence
of a measure term, 
\begin{equation}
S = \alpha N_0 - \beta N_D + \mu M. \label{action2}
\end{equation}
Recent work suggests that the new coupling provides
an expanded phase diagram in which there are lines of first order transitions
in both three and four dimensions \cite{ckrnew}.  A possible phase diagram is
given in fig. \ref{phase}.  The question is: is there a finite value of
$\mu^*$?  If so, the endpoint is presumably a second order phase transition
at which a continuum limit can be taken.
The nearly second order nature of the transition in four dimensions suggests
that $\mu^* \approx 0$ in that case.  In three dimensions studies of the
specific heat suggest $\mu^* \approx -{1 \over 2}$ \cite{ckrnew}.
As a final illustration of the the node deletion scheme, the renormalization
group flows are determined for $\mu = -1.$
Results are shown in fig. \ref{wmeas}.  Typical errors, as a function of the
initial volume, are 0.1, 0.4, 0.6, 1, and 2 for the blocked volume and 0.04,
0.1, 0.1, 0.2, and 0.2 for $<M>$.  All of the flows, for several very large
couplings ($\alpha = 5.9$, $6.4$, and $10.0$), where one would expect to find
the smooth phase, flow toward small node densities.  The largest value of
$\alpha$ at the largest blocking level does begin to move to the left.
This is not necessarily incompatible with a scenario of flows beyond the
end of a first order line, but it
may be that as $\mu$ is made increasingly negative it is necessary to go
to larger volumes to see the transition.  In this case $\mu^*$ would be less
than $-1$.

If critical endpoints of the first order lines actually exist, it will take
more work to nail down their positions.

\acknowledgements
I thank Simon Catterall, John Kogut, and Gudmar Thorleifsson for helpful
discussions.  This work was supported in part by NSF Grant PHY-9503371.

\begin{figure}
\centering
\leavevmode
\epsfxsize=2.0in \epsfbox{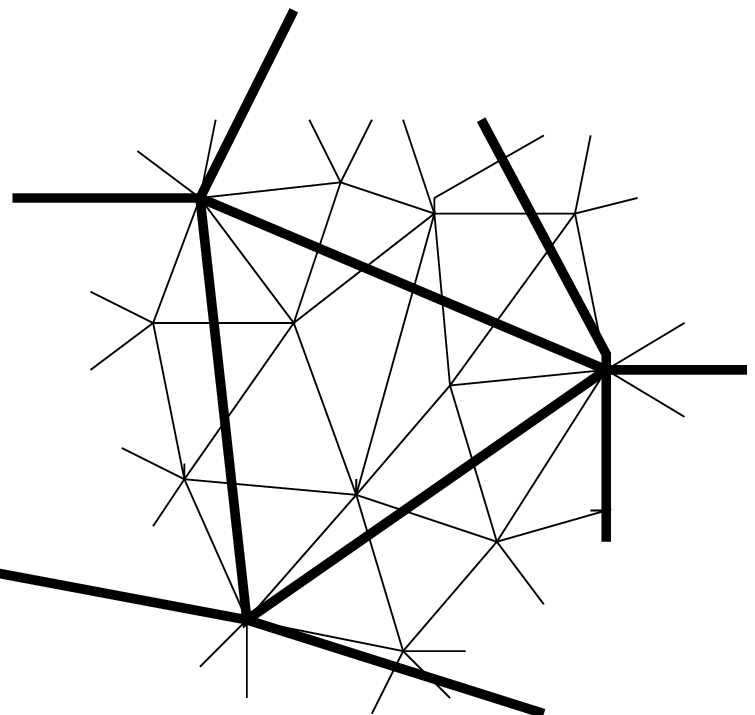}
\caption{ Bare and block triangulations. }
\label{block}
\end{figure}

\begin{figure}
\centering
\leavevmode
\epsfxsize=7.0in \epsfbox{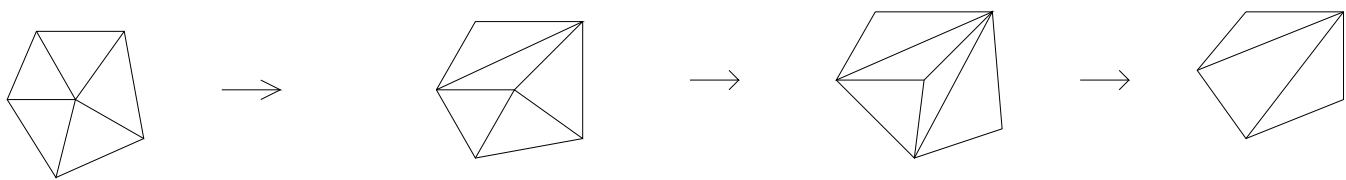}
\caption{ Node deletion in two dimensions. }
\label{delete}
\end{figure}

\begin{figure}
\centering
\epsfxsize=4.7in \epsfbox{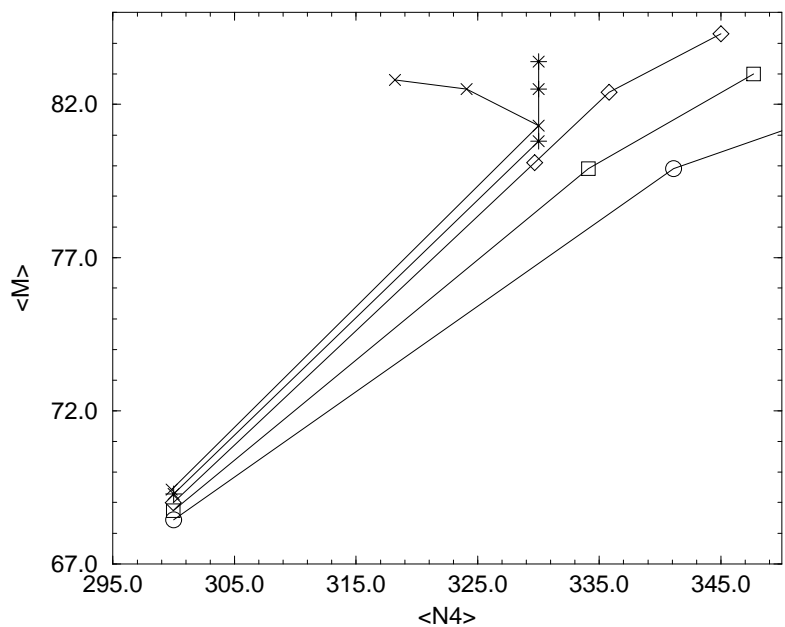}
\caption{Renormalization group flows in the four-dimensional dynamical
         triangulation model near the transition.}
\label{fourd}
\end{figure}

\begin{figure}
\centering
\epsfxsize=4.7in \epsfbox{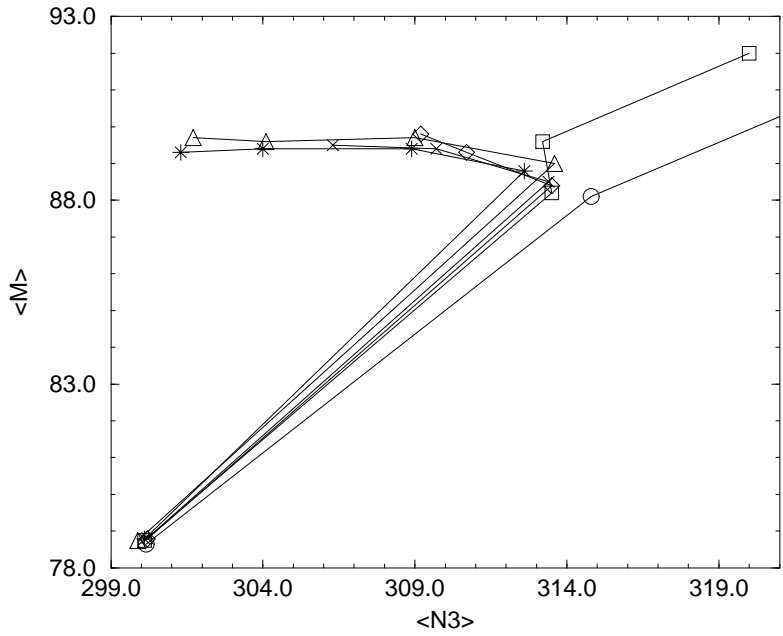}
\caption{Renormalization group flows in the three-dimensional dynamical
         triangulation model near its transition.}
\label{threed}
\end{figure}

\begin{figure}
\centering
\leavevmode
\epsfxsize=5.0in \epsfbox{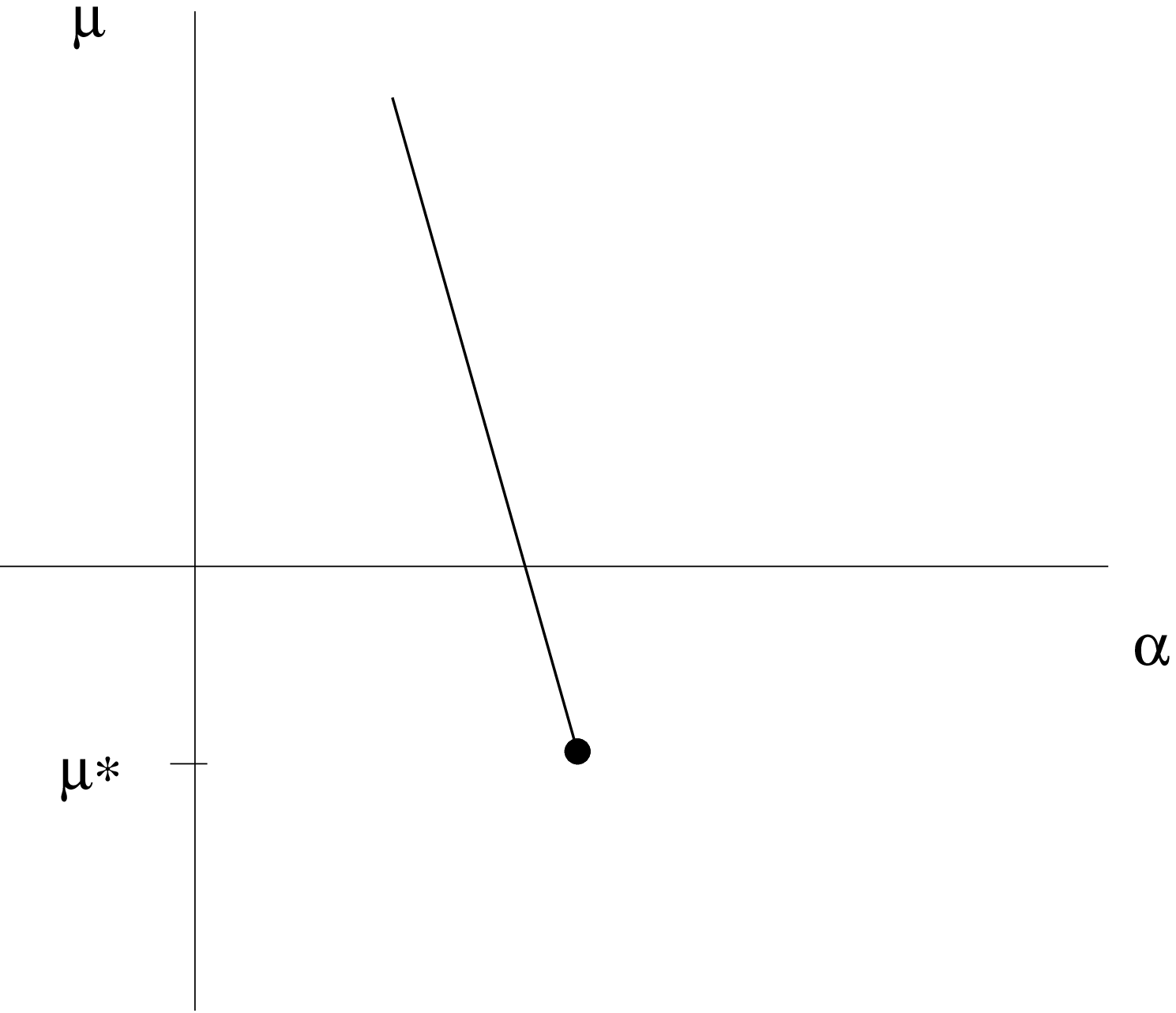}
\caption{ A possible phase diagram when the action includes a measure term. }
\label{phase}
\end{figure}

\begin{figure}
\centering
\epsfxsize=4.7in \epsfbox{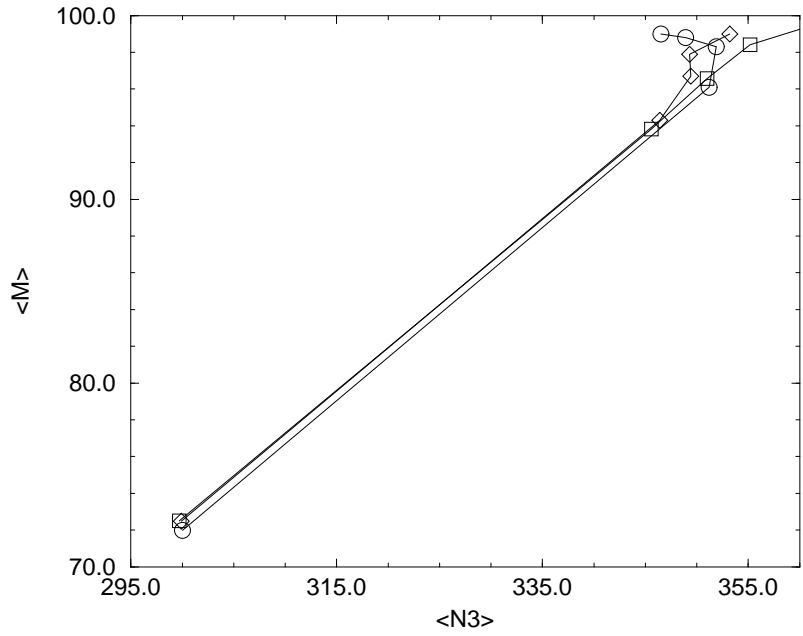}
\caption{Renormalization group flows in the three-dimensional dynamical
         triangulation model with a measure term included.}
\label{wmeas}
\end{figure}

\end{document}